\documentstyle[multicol,aps,psfig,prl]{revtex}

\begin{document}
\title{Pulse confinement in optical fibers with random
dispersion}
\author{M. Chertkov, I. Gabitov, J. Moeser}
\address{Theoretical Division, LANL, Los Alamos, NM
87545}
\date{\today}

\maketitle

\begin{abstract}
Short range  correlated uniform  noise in the  dispersion coefficient,
inherent  in many  types of  optical fibers,  broadens  and eventually
destroys  all  initially   ultra-short  pulses.   However,  under  the
constraint that the integral of the random component of the dispersion
coefficient   is   set   to   zero,   or   pinned,   periodically   or
quasi-periodically  along   the  fiber,   the  nature  of   the  pulse
propagation  changes dramatically.   For the  case that  randomness is
added  to  constant   positive  dispersion,  the  pinning  restriction
significantly reduces pulse broadening.  If the randomness is added to
piecewise constant  periodic dispersion, the pinning  may even provide
probability  distributions of  pulse parameters  that  are numerically
indistinguishable  from the  statistically steady  case.   The pinning
method  can be  used to  both  manufacture better  fibers and  upgrade
existing fiber links.
\end{abstract}



\begin{multicols}{2}

\bigskip  The  effect  of   random  perturbations  in  optical  fibers
increasingly  attracts attention  as  the demand  for  the quality  of
transmission  grows  daily~  \cite{Recent  reviews}.   The  impact  of
randomness on signal transmission in  a single mode fiber is negative;
it   causes  degradation   of   the  signal   and  lowers   transition
capabilities~\cite{Multimode}.       In      particular,     amplifier
noise~\cite{Gordon-Hauss},  and  noise   in  the  fiber  birefringence
(double  refraction)~\cite{PMD} lead  to  random shifts  in the  pulse
position (timing jitter) and  to pulse broadening, respectively.  Both
effects eventually  cause a  destruction of bit-patterns  and eventual
increase  of the  Bit-Error-Rate (BER),  the most  important parameter
describing performance in fiber communications systems~\cite{97A}.

In the  present paper,  we consider the  effect of  random dispersion,
which  is, for  ultrashort  pulses, potentially  as  dangerous as  the
aforementioned effects.   We, however, propose a  way to significantly
reduce the pulse deterioration and eventually reduce the BER caused by
the  noise  in  dispersion  by  using passive  (independent  of  pulse
properties)  periodic control  of  the accumulated  dispersion of  the
fiber  link. Furthermore,  the method  may even  provide statistically
steady propagation of the pulse along the fiber.

Chromatic dispersion  is an important  characteristic of a  medium and
can significantly degrade the  integrity of wave packets. In practice,
chromatic dispersion  is not uniformly distributed  and often exhibits
random  variations  in  space  and  time.  On  the  other  hand,  wave
propagation through  the medium is  usually much faster  than temporal
variations  of  the  chromatic  dispersion.  Therefore,  these  random
variations  can be  treated as  ``spatial'' multiplicative  noise that
does not change in time. This multiplicative noise is conservative and
the  wave  energy  remains  constant during  propagation  through  the
medium.   Recently,  high precision  measurements  of fiber  chromatic
dispersion as  function of a fiber  length experimentally demonstrated
the significance of the dispersion randomness ~\cite{96MMN,98GM}.

The overall  chromatic dispersion in  an optical fiber comes  from two
sources.  The first source is  the medium itself. The second source is
the specific geometry of the waveguide profile. Material dispersion in
the  optical  fiber  is   a  relatively  stable  parameter,  uniformly
distributed  along the  fiber.  However,  waveguide dispersion  is not
nearly as  stable.  Existing technology does not  yet provide accurate
control of  the waveguide geometry of modern  fibers, where dependence
of dispersion on wavelength is complex. As a result, the magnitudes of
random variations of fiber chromatic dispersion are typically the same
as,  or   in  some   cases  even  greater   than  that  of   the  mean
dispersion~\cite{96MMN,98GM}.

In the short-wavelength regime,  a universal description of the signal
envelope in the reference frame  moving with the packet group velocity
is  given by  the  nonlinear Schr\"{o}dinger  equation  (NLS) for  the
complex scalar field, $\psi (z;t)$, see for example \cite{97A},
\begin{equation}
-i\partial _{z}\psi =d(z)\partial _{t}^{2}\psi +2\psi ^{2}\bar{\psi}.
\label{disp}
\end{equation}
The    equation    is    written    in    the    dimensionless    form
\cite{Param}. Variations in the  medium enter this description through
the
dispersion coefficient $%
d(z)=d_{\det }(z)+\xi (z)$, which is decomposed into its deterministic
part, $d_{\det  }(z)$, and a random  part, $\xi (z)$. Here  $z$ is the
position along  the fiber and $t$  is the retarded  time.  The initial
profile $\psi  (0;t)$ is localized  in $t$. We consider  two different
models  of deterministic  dispersion, both  of which  are  standard in
fiber  optic  communications.   Model  (A)  is the  case  of  constant
dispersion, $d_{\det }=d_{0}$. In  the absence of noise ($\xi (z)=0$),
$\psi _{0}(z;t)=a\exp \left[
izd_{0}/b^{2}\right] sech%
\left[  t/b\right]  $,  where  $a^{2}b^{2}=d_{0}$  ($a$  is  the  peak
amplitude, $b$  is the pulse width),  is an exact  soliton solution of
(\ref{disp}). The  existence of the soliton \cite{72ZS}  is the result
of a dynamic equilibrium  between dispersion and nonlinearity: the two
spatial
scales, nonlinearity $%
z_{NL}=1/a^{2}$,  and dispersion $z_{d}=b^{2}/d_{0}$,  coincide. Model
(B)  is the case  of dispersion  management (DM),  $d_{\det }=d_{0}\pm
d_{DM}$ \cite {80LKC}. Here dispersion is piecewise constant: positive
and   negative    spans   alternate   with    period   $z_{DM}$,   and
$0<d_{0}<d_{DM}$.  There is no exact  solution for the pure (no noise)
Model (B), but theoretical  evidence, confirmed by extensive numerical
studies  and  experimental  results,  indicates  the  existence  of  a
breathing  solution   (DM  soliton)  with  a   nearly  Gaussian  shape
\cite{96GT,96SKDBB,99TSSM,99GGDBMGHFG}. The localized solution here is
again due  to the  interplay of dispersion  and nonlinearity.   In the
presence  of a  periodic  dispersion map,  however,  the (DM)  soliton
acquires  an important  characteristic, quadratic  phase  (chirp).  In
contrast to conventional soliton  solutions, DM solitons can exist for
zero (or even negative) values of average dispersion.

Approximate scale  characteristics of the dispersion  noise present in
real     fibers     can     be     extracted     from     experimental
results~\cite{96MMN,98GM}. These results  show that the smallest scale
of noticeable  change in the  dispersion value is  approximately $\sim
1-2  km$. For  constant  dispersion fibers  (model  A), the  amplifier
spacing is $\sim  50-60 km$, and for dispersion  managed fibers (model
B), the  period of a typical  dispersion map is also  $\sim 50-60 km$.
These scales  are much longer  than that of the  dispersion variation,
justifying the  idealized consideration of  delta-correlated noise for
both  models.  Previously,  the  stability of  initial  pulses in  the
presence  of the  short-range-correlated Gaussian  uniform  noise $\xi
_{u}$ with zero mean,
$\left\langle  \xi  _{u}(z_{1})\xi  _{u}(z_{2})\right\rangle  =D\delta
\left(
z_{1}-z_{2}\right)$, 
was studied  for both  models (A) and  (B).  For model  (A), adiabatic
theory, valid if the noise  is weak, shows dynamical broadening of the
pulse and its eventual destruction~\cite{98ACS}.  The pulse-broadening
effects of  $\xi _{u}$  on model (B)  were studied numerically  and by
means of a variational approach~\cite{98ACS,00AB}.

The problem  can also be addressed  in the limit of  strong noise.  On
short scales  the nonlinearity is weak and  propagation is essentially
linear : only  the phase of the pulse is changed  by a rapidly varying
dispersion,  while its  frequency  spectrum is  not.   In this  weakly
nonlinear case  the following change of variables  is suggested, $\psi
\left(  z;t\right)  =\int_{-\infty }^{\infty  }d\omega  \exp \left[  -
i\left(
\omega t+\omega ^{2} %
\left[   \int_{0}^{z}\left(  d(z^{\prime   })-d_{0}\right)  dz^{\prime
}\right] \right) \right] \psi  _{\omega }(z),$ where $\psi_\omega (z)$
describes the pulse evolution on  longer scales.  The equation for the
noise average of the  slow filed $ \varphi _{\omega}=\left\langle \psi
_{\omega }(z)\right\rangle $ derived from (\ref{disp}) is
\begin{eqnarray} & &
\left(  -i\partial  _{z}\!+\!id_{0}\omega ^{2}\right)\varphi  _{\omega
}\!=\!2\int  d\omega_{1,2,3}   \delta  \left(  \omega  _{1}\!+\!\omega
_{2}\!-\!\omega  _{3}\!-\!\omega   \right)  \nonumber  \\   &&  \times
\exp\left[  -i\Delta   \int\limits_{0}^{z}dz^{\prime  }\left(  d_{\det
}-d_{0}\right)\right]\    e^{-\Delta   ^{2}Dz_{\ast    }/2}\   \varphi
_{1}\varphi _{2}\bar{\varphi}_{3}.  \label{kernel}
\end{eqnarray}
Here,   $\Delta    \equiv   \omega   _{1}^{2}+\omega   _{2}^{2}-\omega
_{3}^{2}-\omega  ^{2}$, and  the second  exponential on  the  right of
(\ref{kernel}), which we will call the kernel, is the average of $\exp
\left[ -i\Delta \int_{0}^{z}\xi \left( z^{\prime }\right) dz^{\prime
}\right] $. $z_{\ast}=z$%
,   thereby   setting  the   correlation   scale,  $z_{\xi   }=[\Delta
^{2}D]^{-1}\sim   \lbrack   \omega   ^{4}D]^{-1}\sim  b^{4}/D$.    The
approximation  is  justified,  i.e.    the  nonlinearity  is  weak  if
$z_{\xi}\ll  z_{NL}$.  The exponential  decay of  the kernel  with $z$
disrupts the balance between nonlinearity and dispersion necessary for
steady pulse propagation.

The grim analytical conclusion is that the natural noise in dispersion
leads to destruction of initially  localized signal in the extremes of
both weak
and strong
noise.   Numerical study  of  the intermediate  range  shows the  same
effect.

The natural  question is that of  the existence of  {\em an artificial
constraint}  which  may  reduce   or  completely  prevent  this  pulse
broadening. We  demonstrate that such a constraint  does indeed exist,
and can be readily implemented in real fibers. All that is required is
that {\em the accumulated  dispersion, $\int_{0}^{z}dy\xi (y)$, is set
to zero, or pinned,  either periodically or quasi-periodically} with a
period of the  order or less than $z_{\xi  }$.  The resulting Gaussian
nonuniform noise, $\xi _{n}$
with zero mean is described by 
$\left\langle \xi _{n}(y)\xi _{n}(z)\right\rangle
=D\left( \delta (z-y)-%
\frac{1}{l_{j+1}-l_{j}}\right) $ if
$y$ and $z$ belong to the  same segment bounded by an adjacent pair of
pinning points.
Otherwise there are no correlations.

Consider the  effect of nonuniform  noise, $\xi=\xi_n$, in  the weakly
nonlinear    case.    $z_\ast$    in   (\ref{kernel})    is   $z\left(
l_{j+1}-z\right)  /\left( l_{j+1}-l_{j}\right)  $  with $z\in  \lbrack
l_{j};l_{j+1}]$.  Thus  the decay of  the kernel in  (\ref{kernel}) is
replaced  by oscillatory  (or  quasi-oscillatory, when  the period  is
fluctuating)  behavior. In the  nonuniform case,  additional averaging
over  a  single  pinning  leg  reduces  the  $z$-dependent  kernel  of
(\ref{kernel})  to  $\exp  \left[  -\Delta  ^{2}Dl/4\right]  \sqrt{\pi
/\left[    \Delta   ^{2}Dl\right]    }\mbox{Erfi}\left[   \sqrt{\Delta
^{2}Dl}/4\right] $.  In Model  (B), independent averaging of the first
exponential  on the  right-hand side  of (\ref{kernel})  over $z_{DM}$
(see
also~\cite{96GT,96SKDBB}%
) replaces it by $2\sin \left[ \Delta z_{DM}/4\right]
/\left[ \Delta z_{DM}%
\right] $. The  overall result of the averaging  procedure in the case
of $\xi =\xi _{n}$ is the emergence of a nonlinear term which
does not vanish as $%
z\rightarrow \infty $, in contrast  to the case $\xi =\xi _{u}$, where
the nonlinearity dies away with $z\rightarrow \infty $.
One also concludes that $%
z_\xi$ sets the critical pinning period: the pinning
is efficient only if $%
l=l_{i+1}-l_i\lesssim z_\xi$. It is possible to show that the averaged
equation  for $\xi  =\xi  _{n}$ has  a steady  solution~\cite{00CGMb}.
Notice also  that the averaged kernel decays  exponentially as $\Delta
^{2}\rightarrow \infty $. This is much faster than the algebraic decay
obtained  from  averaging  over  the  dispersion map  period  in  pure
dispersion   management.     Therefore,   the   model   of~\cite{99ZM}
corresponding to  a very narrow  type of nonlinear kernel  (in $\omega
$-space) is  better suited for the  case considered here  than for the
case of  pure DM for which  it was originally  proposed.  The averaged
equation approach  is {\it  \'{a} priori} applicable  for a  large but
finite $z$,  so the emergence of  a steady localized  solution for the
averaged equation does not mean  that the original problem possesses a
steady state.  In the  direct numerical simulation of (\ref{disp}), to
which we now  switch our attention, the effect  is seen as essentially
limiting the pulse broadening.

We perform  numerical investigations of  both models (A) and  (B) with
$\xi =\xi  _{u}$ and  $\xi=\xi_{n}$ in intermediate  case, $z_{NL}\sim
z_\xi$.
Fourier split-step scheme with
$2^{13}$ temporal Fourier modes and periodic conditions imposed on the
boundaries of the domain $t\in [-180,180]$ is implemented.
The spatial step is $%
z_{step}=0.01$ and  the numerical  convergence was checked  by varying
the  size of the  periodic box  and number  of the  Fourier harmonics.
Parameters for the  initial signal were chosen to  be $d_{0}=1$, $a=1$
in
model (A), and $%
d_{0}=0.15$,                 $d_{DM}=0.1$,                 $z_{DM}=1$,
$|\psi(0;t)|=0.79\exp(-t^2/2.6)$ in model (B).  The setup in model (B)
is  borrowed  from~\cite{99TSSM}  and  corresponds  to  experimentally
available   DM  fibers.   Gaussian   zero-mean  noise   correlated  at
$z_{n}=0.1$ with amplitude $d_{n}=1$
models the $\delta $%
-correlated uniform noise with $D=d_{n}^{2}z_{n}=0.1$.  The nonuniform
noise  is  constructed  from   the  uniform  noise  by  the  following
subtraction at every pinning leg: $\xi _{n}(z)=\xi
_{u}(z)-\frac{1}{l_{j+1}-l_{j}}%
\int_{l_{j}}^{l_{j+1}}dy\xi (y)$. Pinning  strategies of two types are
considered: strictly periodic,  $l_{j+1}-l_{j}=l$, where $l$ is fixed;
and  quasi-periodic, $l_{j+1}-l_{j}=l(1+\eta  )$, where  $\eta $  is a
random number  uniformly distributed between $\pm  1/2$.  The averaged
(or otherwise strict) pinning period  for the nonuniform case is taken
to be $1  $, $5$ or $10$. The simulation runs  until the pulse arrives
at  $z=95$.  Statistics  were collected  for $10^2-10^{3}$,  and  in a
special case $10^{4}$, realizations.

\begin{figure}[tbp]
\psfig{file=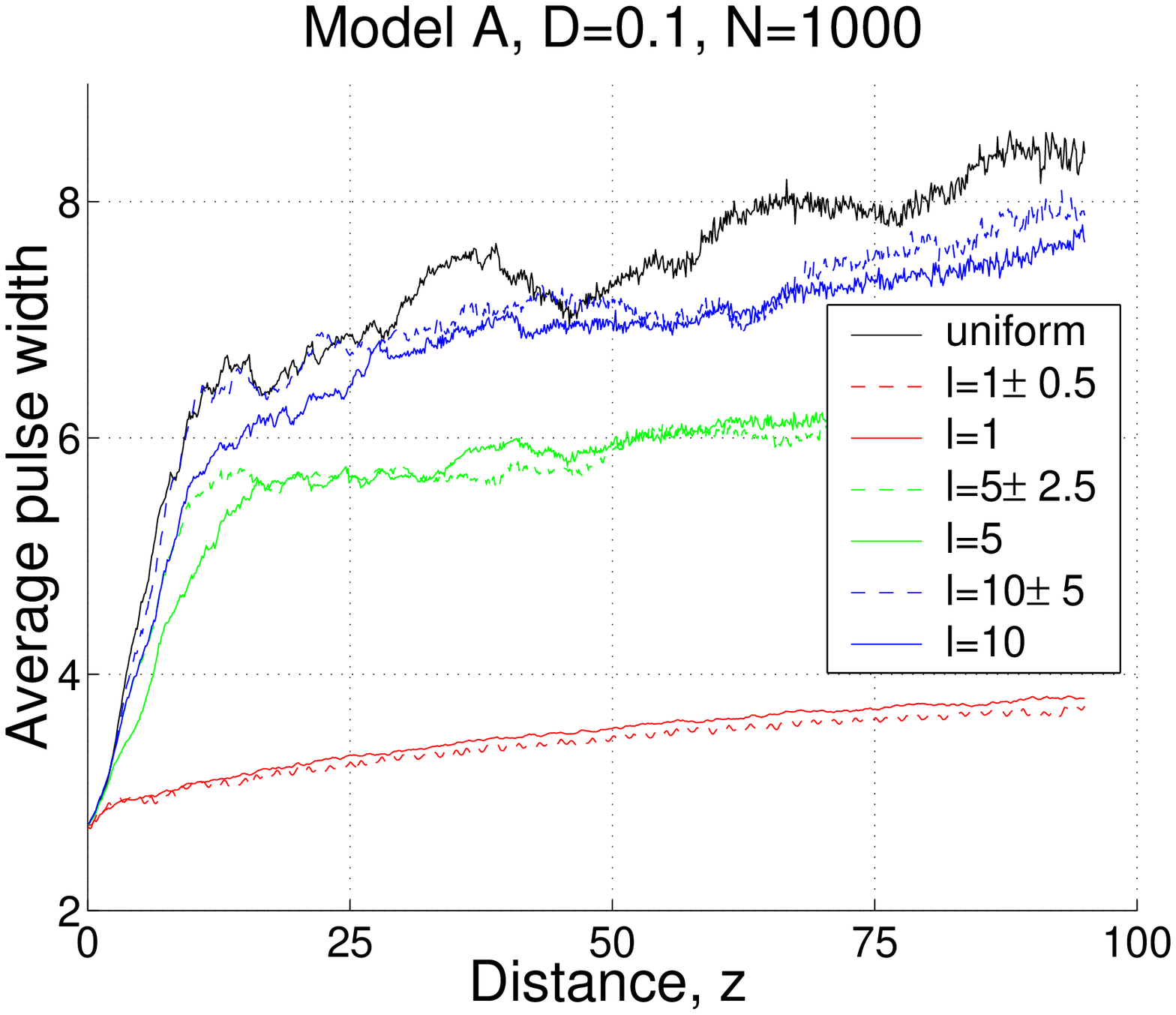,width=8cm,height=5.5cm}
\psfig{file=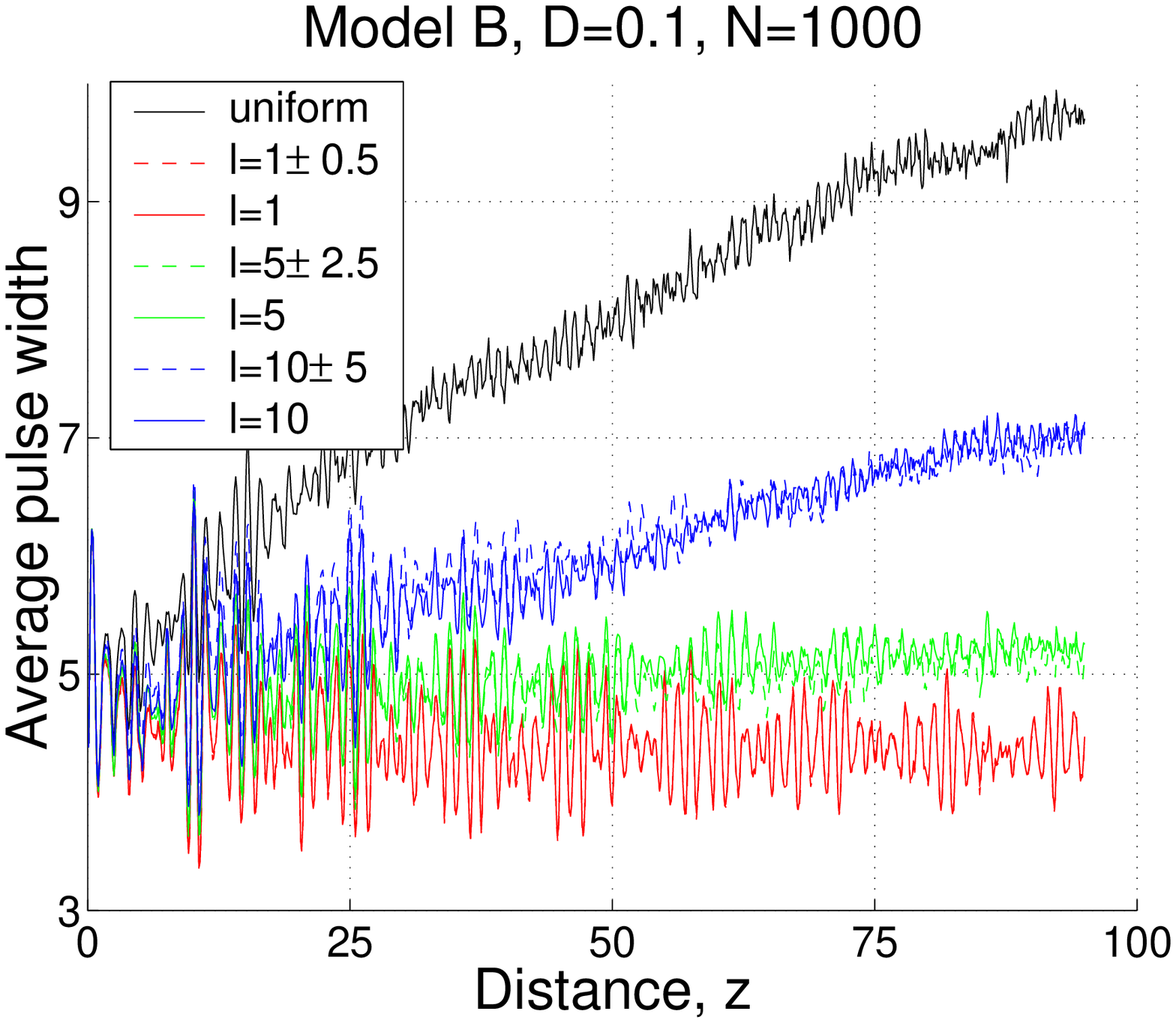,width=8cm,height=5.5cm}
\end{figure}

The averaged pulse-width  (full width at half maximum  amplitude) as a
function  of $z$  is shown  in Figure  1, see  also  \cite{Movie}. The
capital letter  subscript of  the figures corresponds  to the  type of
model, (A) or (B). Solid black, red, green, and blue represent uniform
noise,   and   nonuniform  noise   with   pinning  period   $l=1,5,10$
respectively.  The  quasi-periodic curves are  dashed and of  the same
color as the respective periodic ones.

For Model(A), all types  of nonuniform noise demonstrate a significant
reduction  in the  rate of  pulse  broadening when  compared with  the
uniform  case.  The  individual configurations  that  degrade (through
pulse splitting,  etc.) in the  uniform case maintain  pulse integrity
when each  type of nonuniform compensation is  applied. The dependence
on the  pinning period is monotonic:  the peak amplitude  of the pulse
decays faster as the  pinning period increases. The difference between
periodic and respective quasi-periodic cases is minor, with a slightly
better  confinement   observed  for  the   quasi-periodic  case.   The
destruction of the pulse is also accompanied by emission of continuous
radiation by the  soliton. The radiation is clearly  seen in the movie
made for individual runs~\cite{Movie}.  Once the radiation reaches the
boundaries of  the box, it reflects  and starts to  interfere with the
still localized  solution.  The latter shows  up in the  change of the
averaged-width   behavior  at  larger   distances,  $z\sim   20$.   In
principle, one could introduce absorbing boundary conditions to enable
longer-time simulations. However,  these conditions are artificial, as
in  real  data streams  pulses  are  not  isolated. Each  pulse  emits
continuous  radiation  which  eventually  interacts  with  neighboring
pulses.   In this paper  we purposefully  study self-interaction  of a
pulse and its radiation via reflecting boundary conditions, which is a
simple model for the behavior of a real bit stream. Our results reveal
that after  $z\sim 20$, the  rate of pulse disintegration  is reduced,
suggesting that in  a real bit pattern, pulse  disintegration might be
prevented by radiation absorption from neighboring pulses.

The  effect  of nonuniform  noise  is more  dramatic  in  the case  of
Model(B). For  nonuniform compensation with the  averaged period $l=1$
(and also  less) one observes  a tendency toward  statistically steady
behavior: the  average pulse  width does not  decay (in contrast  to a
decay in  the uniform case),  and the PDF  of the pulse width  (and of
other  variables   characterizing  the  pulse   propagation,  such  as
amplitude)  does not  change shape  with $z$~\cite{D5}.   There  is no
visible emission of  radiation by the localized solution  for any case
of Model (B). We have  also checked that temporal and spatial averages
(for example for the PDF of the pulse width) coincide.  The dependence
on  the  type  of  compensation  (for $l>1$)  is  monotonic,  and  the
difference between the  random and quasi-random cases is  minor, as in
the case of Model(A).

Notice that the tremendous reduction in the pulse decay is achieved by
very minor changes in dispersion (see \cite{Movie}, for the comparable
plots of the dispersion profiles).

Our  analysis  gives  practical  recommendations for  improving  fiber
system  performance  that  is   limited  by  randomness  in  chromatic
dispersion.   The  limitation  originates  from  accumulation  of  the
integral dispersion.  The distance between naturally occurring nearest
zeros of the accumulated dispersion grows with the fiber length ($\sim
\sqrt{z}$  as  it is  shown  in  \cite{00CGMb}).   This growth  causes
asymptotic decay  of the nonlinearity  with $z$ and,  therefore, pulse
degradation.   We have  shown that  the  signal can  be stabilized  by
periodic   or  quasi-periodic  pinning   of  the   accumulated  random
dispersion \cite{different}.  This can  be achieved by first measuring
the mismatch between the nominal  dispersion of the fiber line and the
actual accumulated dispersion, and  second, inserting a small fiber to
compensate this mismatch \cite{00MMGNMGV}.

We are grateful to G.D. Doolen, G. Falkovich, I.
Fatkullin, J. Hesthaven, I.
Kolokolov, P. Lushnikov, P. Mamyshev,  F.G. Omenetto,
H. Rose, T. Schaefer,
Z. Toroczkai, and S. Tretiak  for constructive
comments. M.C. wishes to
acknowledge the support of  J.R. Oppenheimer
fellowship and the hospitality
of ITP Santa  Barbara, where part of this work has
been done. This work was 
supported in part (I.G.) by DOE Contract
W-7-405-ENG-36 and by the  DOE
Program in Applied Mathematical Sciences, KJ-01-01.

\end{multicols}

\begin{references}
\bibitem{Recent reviews}  G. Steinmeyer, D.H. Sutter,
L. Gallmannn, N. 
Matuschek, U. Keller, Science {\bf 286}, 1507 (1999);
G.A. Thomas,  A.A.
Ackerman, P.R. Prunchal, S.L. Cooper, Physics Today
{\bf 53,  9}, 30 (2000); 

\bibitem{Multimode}  Notice, however, that in the
multi-channel  situation
(which will not be discussed here) the opposite to the
 standard expectation
is also possible : namely the presence of random  inter-modes
(channels) scattering
can be used here to improve the  total information
capacity of a multi-mode
optical fiber. See, A.F.  Garito, J. Wang, R. Gao,
Science {\bf 281}, 962
(1998); H. Stuart,  Science {\bf 289}, 281 (2000), and
also discussion of
similar ideas  in the field of wireless
communications: G. J. Foschini, M.J.
Gans,  Wireless Personal Commun. {\bf 6}, 311 (1998);
A.L. Mousstakas,  H.U.
Baranger, L. Balents, A.M. Sengupta, S.H. Simon,
Science {\bf \ 287}, 287
(2000).

\bibitem{Gordon-Hauss}  J.N. Elgin, Phys. Lett. {\bf
110A}, 441 (1985); 
Phys. Rev. E {\bf 47}, 4331 (1993);J.P. Gordon, H.A.
Haus, Opt.  Lett. {\bf %
11}, 665 (1986); G. Falkovich, I. Kolokolov, V.
Lebedev,  S. Turitsyn,
preprint 04/2000, 
http://xyz.lanl.gov/abs/nlin.CD/0004001.

\bibitem{PMD}  C.D. Poole, Opt. Lett. {\bf 13}, 687
(1988); {\bf 14},  523
(1989); C.D. Poole, J.H. Winters, and J.A. Nagel, Opt.
Lett. {\bf 16}, 372
(1991); N. Gisin, Opt. Comm. {\bf 86}, 371-373 (1991);
 P.K. Wai,
C.R.Menyak, H.H. Chen, Opt. Lett. {\bf 16}, 1231
(1991).

\bibitem{97A}  G.P. Agrawal, Fiber-optic communication
systems, New  York,
Wiley, 1997.

\bibitem{96MMN}  L.F. Mollenauer, P.V. Mamyshev, M.J.
Neubelt, 
Opt. Lett. {\bf 21}, 1724 (1996).

\bibitem{98GM}  J. Gripp, L.F. Mollenauer, 
Opt. Lett. {\bf 23}, 1603 (1998).

\bibitem{Param}  All parameters are presented here in
dimensionless units
which transform to real-world fiber units according to
the following rules.
The envelope of the electric field is in the form
$\psi =E/\sqrt{P_{0}}$,
where $P_{0}$ is the peak pulse power. The propagation
variable is $%
z=x(\alpha P_{0}/2)$, where $x$ is distance along the
fiber and $\alpha $ is
the Kerr nonlinearity coefficient. The Kerr
coefficient can be expressed in
terms of other fiber parameters, $\alpha =2\pi
n_{2}/(\lambda S_{eff})$,
where $n_{2}$ is the nonlinear component of fiber
refractive index, $\lambda 
$ is operating wavelength, and $S_{eff}$ is an
effective core area of the
fiber. The spatial coordinate is $t=\tau /\tau _{0}$,
where $\tau $ is in
the reference frame of the group velocity and $\tau
_{0}$ is the
characteristic pulse width. The dispersion coefficient
is $d=2\beta
_{2}/(\alpha P_{0}\tau _{0}^{2})$, where $\beta _{2}$
is the second order
dispersion parameter. The typical parameters for
dispersion shifted fiber
are: $\lambda =1550$nm , $\tau _{0}=7.01$ps, $x=50$km,
$P_{0}=$ $4$mW, $%
\beta _{2}$ $=2$ ps$^{2}/$km, $\alpha =10$
Wt$^{-1}$km$^{-1}$.

\bibitem{72ZS}  V. E. Zakharov, A.B. Shabat, Sov.
Phys. JETP {\bf 34},  62
(1972).

\bibitem{80LKC}  C. Lin, H. Kogelnik, L.G. Cohen, Opt.
Lett. {\bf 5},  476
(1980).




\bibitem{96GT}  I. Gabitov, S.K. Turitsyn, Opt. Lett.
{\bf 21}, 327  (1996);
JETP Lett. {\bf 63}, 861 (1996).

\bibitem{96SKDBB}  N. Smith, F. M. Knox, N.J. Doran,
K.J. Blow, I.  Bennion, 
Electron. Lett. {\bf 32}, 54 (1996).

\bibitem{99TSSM}  S.K. Turitsyn, T. Schafer, K.H.
Spatschek, V.K. 
Mezentsev, 
Opt. Comm. {\bf 163}, 122 (1999).

\bibitem{99GGDBMGHFG}  D.Le. Guen, et al.,
Post Deadline Paper, OFC/IOOC'99, San
Diego, CA, USA.

\bibitem{98ACS}  F. Kh. Abdullaev, J.G. Caputo, M.P.
Sorensen, in New 
trends in Optical Soliton Transmission Systems, A.
Hasegawa, Eds.  (Kluwer,
Dordrecht, 1998); F.Kh. Abdullaev, J. C. Bronski and
G.  Papanicolaou,
Physica {\bf D}135, 369 (2000).

\bibitem{00AB}  F. Kh. Abdullaev, B. B. Baizakov, 
Opt. Lett. {\bf 25}, 93 (2000).

\bibitem{00CGMb}  M. Chertkov, I. Gabitov, J. Moeser,
Z. Toroczkai, 
unpublished.

\bibitem{99ZM}  V.E. Zakharov, S.V. Manakov, JETP
Lett. {\bf 70}, 578 
(1999).

\bibitem{Movie}   Movies of single realization
dynamics in $z$, comparative
plot of the dispersion profile with and without
compensation, and more
figures characterizing statistics of the pulse
propagation (in the various
cases considered) are available at
http:/cnls.lanl.gov/\symbol{126}chertkov/Fiber.

\bibitem{D5}  Notice that for the case with the same
pinning period $l=1$
but  greater $D=2.5$, a minor, but still observable
degradation of pulse 
occurred. This is consistent with the statement
concerning the  efficiency
of the pinning made in the text: the greater $D$, the 
lower the critical
pinning period.

\bibitem{different}  Real fiber spans, usually each of
length $0.5-2$ $z_{NL}
$, are not homogeneous, as they are often produced at
different  times by
different companies.

\bibitem{00MMGNMGV}  After the work was completed we
learned about the 
paper of L.F.Mollenauer, et al. [ Opt. Lett. {\bf 25},
704 (2000)],  where
long haul transmission experiments on fibers
built from  spans of
different types are described. It was shown there that
 periodic
compensation of the overall dispersion to the fixed 
residual value
(achieved via insertion of an extra span) optimizes 
propagation of pulses.
These results are consistent with, and also  give an
experimental support
to, the present proposed pinning  method.
\end{references}
\end{document}